\documentstyle[10pt,editedvolume,numreferences]{crckapb}

\begin{opening}
\title{Structure of  light.}
\author{Rohini M. Godbole}
\institute{Physics Dept., Bombay Univ., Vidyanagari, Santa Cruz (East),
Bombay 400098, India}
\end{opening}
\runningtitle{Structure of Light}
\begin{document}
\setlength{\textheight}{19cm}
\setlength{\textwidth}{12.5cm}
\def\be{\begin{equation}}
\def\ee{\end{equation}}
\newcommand{\een}{\end{subequations}}
\newcommand{\ben}{\begin{subequations}}
\def\bea{\begin{eqnarray}}
\def\eea{\end{eqnarray}}
\def\ptmin{\mbox{$ p_{T,min} $}}
\def\gamgam{\mbox{$\gamma \gamma $}}
\def\qvph{\mbox{${ \vec f_{q/{\gamma}}} $}}
\def\qph{\mbox{${f_{q/\gamma} } $}}
\def\vqxqsq{\mbox{$ \vec f_{q/{\gamma}} (x,Q^2)  $}}
\def\vqisq{\mbox{${ f_{q_{i}/\gamma} (x,Q^2) }$}}
\def\vqisqm{\mbox{$ f_{q_{i}/{\gamma}} (x,Q^2)$ }}
\def\gamg{${ \gamma g }$}
\def\eplem{\mbox{$e^+ e^- $}}
\def\rts{\mbox{$ \sqrt{s} $}}
\def\qsq{\mbox{$Q^2$}}
\def\qqbar{\mbox{$q \bar{q}$}}
\def\pbarp{$ \bar p p$}
\def\alphas{\mbox{$ \alpha_s$ }}
\def\gamp{${ \gamma p}$}
\def\totwo{$2 \rightarrow 2$}
\def\glph{\mbox{${ f_{G/{\gamma}}(x,Q^2)}$}}
\def\fgme{\mbox{$ f_{\gamma/e}$ }}
\def\sigh{ \hat \sigma}
\def\alpas{\mbox{${ \alpha_s}$}}
\def\f2gam{\mbox{$ F_2^\gamma $}}
\renewcommand{\thefootnote}{\fnsymbol{footnote}}
\begin{abstract}
In this talk I  briefly explain the concept of the structure function of a
photon (the best known boson). Then I review some of the current experimental
evidence which confirms the existence of `strong' interactions of photon
suggested by this idea.  I end by pointing out how the photon `structure' has
important implications for the interactions of high energy photons and hence
for the design of the next generation of the high energy $e^+e^-$ (linear)
colliders which are absolutely essential for locating the missing links in our
knowledge of fundamental particles and interactions among them.

\end{abstract}

\section{Introduction}
This symposium is being held to celebrate the birth centenary of
S.N. Bose after whom half of  the elementary particles are named
, i.e., the particle with integral spin: bosons. In this talk I want to
discuss an interesting feature of interactions of the one boson best known to
the particle  physics community {\it viz.} the photon. In spite of the fact
that the theory of interaction of photons with electrons is the best
formulated, most studied and best tested theory, interactions of photons with
matter continue to exhibit interesting features which give further insight into
the question of fundamental constituents of matter and interactions among them.
At high energies, the measured cross-sections in processes involving
photons seem to imply that the elementary, point-like photons behave at high
energies like strongly interacting particles  (hadrons) which are bound states
of  more fundamental quarks and antiquarks. In this talk I discuss
the issue of this `structure' of photons and the implications of this for the
design and planning of the next generation high energy $e^+e^-$ colliders.

\section{Standard Model and Supercolliders}
 To understand the  concept of photon
structure and its implications for the high energy  photon interactions
clearly, let us first summarize briefly our current understanding of the
fundamental constituents of matter and interactions among them:
the standard model(SM). The fundamental (elementary) constituents of
matter are the spin $1/2$ fermions (along with their antiparticles)
summarised in
{
\begin{table}[hbt]
\caption{Fundamental constituents of matter}
\begin{center}
\begin{tabular}{ll}
\hline
Quarks&Leptons\\
\hline
\\
$ \left(
\begin{array}{c}
u \\
d
\end{array}
\right) \,\,
\left(
\begin{array}{c}
c \cr
s
\end{array}
\right)\,\,
\left(
\begin{array}{c}
t\cr
b
\end{array}
\right)\,\, \times 3 \, {\rm colours}$
& $
\left(
\begin{array}{c}
\nu_e \\
e^-
\end{array}
\right)\,\,
\left(
\begin{array}{c}
\nu_{\mu}\\
\mu^-
\end{array}
\right)\,\,
\left(
\begin{array}{c}
\nu_{\tau}\\
\tau^-
\end{array}
\right)
\,\,
$
\\
\\
\hline
\end{tabular}
\end{center}
\end{table}
}
Table 1. There exist four fundamental  interactions among
these matter particles, out of which only the strong, electromagnetic and weak
interactions are relevant for this discussion. The corresponding coupling
constants satisfy the hierarchy $g_s >  g_e=e >  G_W$.
All the interactions
among these fundamental constituents can be understood in terms of exchange of
spin 1 (vector) bosons. The theoretical framework for their description is that
of the gauge field theory and the interaction mediating bosons are called gauge
bosons. The  mediators of the electromagnetic and strong interactions (the
photon $\gamma $ and the gluon $g$ ) are massless whereas those corresponding
to
weak--interactions viz. $W^{\pm}/Z$ are massive. The massive nature of the
$W^{\pm}/Z$ would normally spoil the gauge symmetry and hence would come in the
way of a gauge field theoretical  description of the weak--interactions.
However, the ingenious mechanism of spontaneous  breakdown of the gauge
symmetry (SSB), where the symmetry of the vacuum and the lagrangian are
different, makes it possible to have such a description. However,
this description requires existence of one more elementary spin 0 boson,
the Higgs scalar, in addition to  the twelve
gauge bosons and the fermions listed in Table 1. This mechanism
provides a rather neat way of giving masses to the fermions as well. At
present all the features of this picture (SM) have been verified to a great
accuracy except the existence of the Higgs boson. It is the quest for this,
still missing, member of the SM that mainly prompts the planning of the next
generation  of high energy colliders. It can be argued on very general grounds
that experiments around an energy scale 1 TeV, aught to either find this Higgs
scalar or confirm that the solution to the basic problem of mass generation for
the $W^{\pm}/Z$ and the fermions lies somewhere else other than in the SSB
mechanism and give us hints about the possible mechanism which achieves this.
It should also be mentioned here that as far as the particle physicists are
concerned the SSB mechanism is theoretically the most attractive and  only
truly viable mechanism that exists at present.  The supercolliders, which are
required to have super--high energies and luminosities, that are currently
under discussion are the $p \bar p$ collider (LHC) with a total centre of mass
(c.m.)  energy ($\sqrt{s}$) of $\sim 15 $ TeV $ (1 {\rm TeV} = 1000 {\rm
GeV})$  with a luminosity of $\sim 10^{34}/cm^2/sec$
and an $e^+e^-$ collider with a c.m. energy $\rts \geq 500$ GeV and ${\cal L}
= 10^{33}/cm^2/sec$. A point to note also is that these colliders are
expected to operate at much higher energies and higher luminosities than at the
current colliders : \rts \ = 2 TeV for a  \pbarp\ collider with
${\cal L} \simeq  10^{31} $  at the Tevatron at FNAL and $ \rts \simeq 100$
GeV for an \eplem\ collider with ${\cal L} = 5 \times 10 ^{30}/cm^2/sec$
at LEP at CERN. Hence potentailly new phenomena might occur at
these supercolliders.

\section{Structure of matter}
The terminology of the `structure ' of a photon is essentially a
short hand way of describing how a high energy photon interacts
with other particles: hadrons and photons. It does not of course
imply that the  $\gamma$ is not an elementary particle.
As an introduction to photon structure
let us briefly understand how one describes structure of matter in general.
According to the currently accepted  picture all the strongly interacting
particles observed in nature (called hadrons) are bound states of quarks,
antiquarks and gluons. The interactions among these hadrons, at high
energies, are described in terms of those between the constituents,
$q,\bar q $ and  $g$, collectively called partons. This is the so called
parton--model picture shown in
\begin{figure}[hbt]
\vspace{4.5cm}
\caption{Parton Model}
\label{pmodel}
\end{figure}
fig. \ref{pmodel}.  This picture is rigorously
proved in the perturbative Quantum--Chromo--Dynamics (pQCD) which is the field
theoretical description of the strong interactions in terms of an $SU(3)$ gauge
theory.  Such a description of high energy processes requires, in addition to
the knowledge of QCD, also the information on the parton content of the
hadrons viz: the parton density functions $f_{p_1/H_1} (x_1)$; the probabilty
of finding a parton $p_1$ in hadron $H_1$ carrying the momentum fraction $x_1$
of the hadron $H_1$. The  functions $f_{p_i/H_i} (x_i)$ can not, as yet, be
computed from first principles in QCD and have to be measured experimentally.
This information is obtained by studying the deep inelastic scattering
(DIS) of high energy leptons of energy E off hadron targets,
\be
e^- + H \rightarrow e^- + X
\label{dis}
\ee
The double differential  cross--section  for the process is a
function of two independent variables $y = \nu /E $  where
$\nu $ is the energy carried by the probing photon in the laboratory
frame, and $x=Q^2 /( 2 M \nu ) $ where M is the proton
mass and $ - Q^2 $ is the invariant mass of the virtual photon
\begin{figure}[hbt]
\vspace{3.5cm}
\caption{Deep Inelastic Scattering for the proton and photon.}
\label{disfig}
\end{figure}
in fig. \ref{disfig} (a) which shows the DIS process for a proton.
In the quark-parton-model (QPM) this double differential cross--section
is given by,
\be
\frac{d^2\sigma^{ep \rightarrow X}}  {dx dy} = \frac {2 \pi\
\alpha^2 \ s}
{Q^4} \times \bigg[(1+(1-y)^2)\ F_2^p (x) - y^2\ F_L^p(x) \bigg],
\label{discs}
\ee
where
\bea
 F_2^p(x) &  = &  \sum_{q} e_q^2 \  x\ f_{q/p}(x); \nonumber \\
 F_L^p(x) & = & F_2^p (x) - 2 x F_1^p(x) \nonumber
\eea
are the two electromagnetic structure functions of the proton and
$ f_{q/p}(x) $  the probability for quark $q$ to carry a momentum
fraction $x$ of the proton and $e_q$ denotes the electromagnetic
charge of quark $q$ in units of the proton charge.
 QCD implies some corrections to the QPM and these give the
structure function $F_2^p $ a \qsq\ dependence which is given by
the evolution equations \cite{glap} predicted in pQCD. The corrections
also change $F_L^p(x,Q^2)$ from its QPM value of zero.  But we will not concern
oursleves with these here.

To measure the structure function of a photon such an
experimental situation is provided at \eplem\ colliders in
$\gamma^* \gamma $ reactions as shown in fig.~\ref{disfig} (b).
Here the virtual photon with invariant mass square $-Q^2 $
probes the structure of the real photon.  The idea that
photons behave like hadrons when interacting with other hadrons
dates back to the early days of strong interaction physics and
is known to us under the name of the Vector Meson Dominance
(VMD) picture. This essentially means that at low 4--momentum
transfer, the interaction of a photon with hadrons is dominated
by the exchange of vector mesons which have the same quantum
numbers as the photon.  If the VMD picture were
the whole story then one would expect that such an experiment
will find
\be
F_2^\gamma \simeq F_2^{\gamma,VMD} \propto F_2^{\rho^0} \simeq F_2^{\pi^0}.
\label{f2VMD}
\ee
Then with increasing \qsq,
the  structure function \f2gam\ will behave just like a hadronic
proton structure function.
However, there is a very important
difference in case of photons, {\it i.e.}, photons possess pointlike
couplings to quarks. This has interesting implications for $\gamma^* \gamma $
interactions as first noted in the framework of the
QPM by Walsh \cite{wal}.  It
essentially means that $\gamma^* \gamma $ scattering in
fig.~\ref{disfig} contains two contributions as shown in
\begin{figure}[hbt]
\vspace{4.0cm}
\caption{Two contributions to \f2gam.}
\label{fgamtwo}
\end{figure}
fig.~\ref{fgamtwo}. The contribution of fig.~\ref{fgamtwo}~(a) can be
estimated by eq.(\ref{f2VMD}), whereas that of fig.~\ref{fgamtwo}~(b)
was calculated in the QPM~\cite{wal}. The dominant contribution  comes from the
kinematical region when the quark in the t and u channel is on mass-shell
and hence  can be calculated only when one considers
quarks with finite masses. The result can  be recast in a form equivalent to
eq. (\ref{discs}):
\bea
{ {d^2\sigma^{e\gamma \rightarrow X}} \over {dx dy}} &= &{ {2
\pi
\alpha^2  s_{e \gamma}} \over {Q^4}} \times
\nonumber \\
& & \bigg[{3 \alpha \over \pi} {\sum_q e_q^4 \bigg\{ (1+(1-y)^2)
\times {\bf [} x (x^2 +(1-x)^2) \times \ln{{W^2} \over {m_q^2}}} \nonumber \\
& & +{ 8x^2 (1-x) -x{\bf ]} - y^2 {\bf [} 4 x^2 (1-x)
{\bf ]} \bigg\}}\bigg],
\label{qpm}
\eea
where $ W^2 = Q^2 (1-x)/ x.$
On comparing  eqs. (\ref{discs}) and (\ref{qpm}) we see that the
factors in square brackets in the above equation  have the
natural interpretation as photon structure functions $ F_2^\gamma $ ,
$ F_L^\gamma $ and one has
\bea
F_2^{\gamma,\rm{pointlike}}(x,Q^2) &= & 3 {\alpha \over \pi}
                            { \sum_q e_q^4 \bigg[ x (x^2 +(1-x)^2)
                             \times
\ln{{W^2} \over {m_q^2}} + 8x^2(1-x) -x \bigg]} \nonumber \\
                       & = & {\sum_q e_q^2\;  x\
                       f_{q/\gamma}^{\rm{pointlike}}(x,Q^2)}.
\label{qpmpred}
\eea
Two points are worth noting: the function $
F_2^{\gamma,\rm{pointlike}}(x,Q^2) $ can be completely calculated in
QED and secondly this contribution to \f2gam\ increases
logarithmically with \qsq. So in this simple `VMD + QPM' picture,
\f2gam\ consists of two parts, $F_2^{\gamma,\rm{pointlike}}$ and $
F_2^{\gamma,\rm{VMD}}$, with distinctly different \qsq\ behaviour and
with the distinction that for one part both the $x$ and the \qsq\
dependence can be calculated completely from first principles.

This QPM prediction received further support when it was shown by
Witten \cite{wit77} that at large \qsq\ and at large $x$, both the $x$
and \qsq\ dependence of the quark and gluon densities in the photon
can be predicted completely even after QCD radiation is included.  An
alternative way of understanding this result is to consider the
evolution equations \cite{dewit} for the quark and gluon densities
inside the photon.  In the `asymptotic' limit of large \qsq\ and large
$x$, the \vqisq\ have the form
\bea
f_{q_i/{\gamma}}^{\rm asymp}(x,Q^2)& \propto & {\alpha \times
\ln \left( {Q^2}
\over {\Lambda^2_{\rm{QCD}}} \right) F_i(x) }\nonumber \\
&\simeq & {{\alpha \over {\alpha_s}} F_i(x)},
\label{asymp}
\eea
where $\rm \Lambda_{QCD} $ is the  QCD scale parameter,
$\alpha_s (Q^2) $  is given in terms of the running strong coupling
constant by $g_s^2 (Q^2) / 4 \pi $ and the $x$
dependence of the $F_i(x)$ is completely calculable. Note here
the factor  $\ln \left( {Q^2} \over {\Lambda^2_{\rm{QCD}}}
\right)$ on the r.h.s.~Measurements \cite{exrev} of the photon
structure function $F_2^{\gamma}$ in $\gamma^* \gamma$ processes
did indeed confirm the basic QCD predictions of the linear rise
of $F_2^\gamma$ with $\ln \left(Q^2\right)$ at large $x$.
This discussion thus means that just like one can `pull' quarks
and gluons out of a proton one can look upon the photon as a
source of partons and that the parton content of the photon
rises with its energy. Physically this means that the photon
splits in a $ q \bar q$ pair and these radiate further gluons
and thus fill up a volume around photon with partons.

The asymptotic solutions discussed above, though very useful to
understand the rise of the photon structure function with \qsq , are
valid only at large $x$ and large \qsq . At small values of $x$ these
solutions diverge, indicating thereby that `hadronic' part of \f2gam\
can not be neglected at small $x$. Hence it is now generally
accepted that for practical purposes, specially if one wants to use the parton
model language for the interactions of high energy photons, it is better
to forego the absolute predictions of \f2gam\ of the asymptotic part,
that are possible in pQCD and use only the prediction of the \qsq\
evolution of the
photon structure function in analogy to the case of the proton
structure function. At present there exist fourteen different
parametrisations of the photon structure function
\cite{comment}. The DIS measurements described above measure
only the quark-parton densities \vqisq\ (for $x > 0.05 $ and $Q^2 <
100-200$ GeV$^2$) directly and \glph\ is only inferred indirectly. As
a result there is considerable uncertainty in the knowledge of
\glph. The different parametrisations differ quite a lot from each
other in the gluon content. It should also be mentioned here, that
these differences reflect the differences in different physical
assumptions in getting \glph\ from the data on \f2gam. So independent
information on \glph\ is welcome.

\section{ Calculation of jet production in \gamgam,
\gamp\ collisions}

One such possibility is the study of jet production in \gamgam\ ,
\gamp\ collisions.
Jet production in \gamgam\ collisions can receive contributions from
three different types of diagram \cite{llewellyn} as shown in
\begin{figure}[hbt]
\vspace{9.5 cm}
\caption{Different contributions to the production of high $p_T$
jets in \eplem\  collisions with the associated topologies.}
\label{ptjts}
\end{figure}
fig.~\ref{ptjts}. The `direct process' of fig.~\ref{ptjts}a is due
to $\gamgam \rightarrow \qqbar\ $  production, present already in the
naive quark-parton model. Fig.~\ref{ptjts}b depicts the case where
only one photon is resolved into its partonic components, which then
interact with the other photon; we call these the `once-resolved'
processes (`1-res' for short).  Finally, fig.~\ref{ptjts}c shows
the situation where both photons are resolved, so that the hard
scattering is a pure QCD \totwo\ process; we call these the
`twice-resolved' contributions (`2-res' for short). It is very
important to note here that every resolved photon will produce a
spectator jet of hadrons with small transverse momentum relative to
the initial photon direction, which for (quasi-~) real photons
coincides with the beam direction. The resolved contributions of
fig.~\ref{ptjts}b and c can therefore be separated if one can
tag on these spectator jets.

The cross-section for jet production in \gamgam\ collisions (the $e^-/e^+$ acts
as the source of `almost' real photons when the $e^-/e^+$ is scattered at very
small angles, and thus \eplem\ collisions can be used to study \gamgam\
collisions) for the
`2-res' processes can be written  as \cite{oldjet,tristanac}
\bea
 {d \sigma \over dp_T}  =  \sum_{p_1,p_2,p_3,p_4}
\int_{z_{1min}} dz_1 \;  f_{\gamma_1/e}(z_1)
\int_{z_{1min}/z_1} dz_2 \; f_{\gamma_2/e}(z_2) \nonumber \\
\int_{z_{2min}/z_2} dx_1  \; f_{p_1/\gamma_1} (x_1)
\int_{x_{1min}/x_1} dx_2 \;  f_{p_2/\gamma_2} (x_2)\nonumber \\
\times {d \hat{\sigma} (p_1 + p_2 \to p_3 + p_4) \over dp_T},
\label{csec}
\eea
where the $ d \hat{\sigma} / dp_T $ are the cross sections for the hard \totwo\
subprocesses,  $  f_{p_j/\gamma_i} (x_j,Q^2)$ , $ f_{\gamma_i/e}(z_i) $
denote parton densities inside the photon and  photon fluxes inside the
electron respectively and $z_{1min} = 4 p_T^2 / s $. For
the `1-res' (direct) processes, one (both) of the parton density functions
$f_{p_i/\gamma_j}(x_i)$ have to be replaced by $ \delta (1-x_i)$,
and the proper hard sub-process cross--sections have to be inserted.  Recall
eq.(\ref{asymp}) for $f_{q_i/\gamma} (x_i, \qsq)$. This relation  makes it
clear that all three classes of diagrams are of the same order
in $\alpha$ and \alpas. The `resolved' events will also have additional
`spectator' jets in the direction of the $\gamma$, i.e, in the direction
of the $e^-/e^+$.
\begin{figure}[hbt]
\vspace{6.5 cm}
\caption{$ d \sigma / d p_T$ at $p_T = 3$ GeV as a
function of $\protect\sqrt{s}$ \protect\cite{tristanac}.}
\label{csrts}
\end{figure}
Fig. \ref{csrts} shows the energy dependence of the cross--section
for the production of two jets with $p_T = 3$ GeV,
as predicted \cite{tristanac} for one of the  parametrizations of \f2gam ,
in the range covered by the  \eplem\ colliders PETRA and TRISTAN .
The cross--section is also well above the background from annihilation
events with hard initial state radiation (dotted curve). More importantly
twice--resolved contribution grows faster than \rts\ with increasing
machine energy and, for this choice of $p_T$, begins to dominate the
cross--sections in the energy range of TRISTAN. The $\gamma$ energies increase
with the \rts\ of the \eplem\ machine. With increasing $\gamma$ energies
increasingly more energy becomes available to the partons participating in the
subprocess, for a fixed $p_T$  or inv. mass of the final state. Hence the
importance of the `resolved' processes increases with increasing energy.
Experimental studies of the jet--production in \gamgam\ collisions
\cite{uehara} at the \eplem\ colliders TRISTAN and LEP, have
confirmed the existence of the `resolved' contributions
\cite{tristanac}.  These studies have even ruled out some of
the very hard parametrisations  of $\glph$ \cite{uehara} as shown in
\begin{figure}[hbt]
\vspace{6 cm}
\caption{$ d \sigma / d p_T$ at  TRISTAN compared with the theoretical
predictions \protect\cite{uehara}.}
\label{uehara}
\end{figure}
fig. \ref{uehara}.

Jet production in $ep$ (or equivalently $\gamma p$ ) collisions also
has two contributions : `direct' and `resolved'. High energy photons
are effectively available at the HERA collider at DESY, in the collision of
a 30 GeV $e$ beam with a 820 GeV $p$ beam. This corresponds to a c.m. energy
$\leq 300 $ GeV , which in turn corresponds to $E_{\gamma} \leq 50$ TeV.
Our calculations \cite{dgprd} showed that here also the  photo-production of
jets is dominated by the `resolved' contributions upto $p_T = 40$ GeV. The
`resolved' contributions are expected to have `spectator' jets in the direction
of the $\gamma$ (i.e. the electron).  This rate also depends strongly on
\glph, \vqisq  and hence can be used to get information about these.
Recent measurements at HERA \cite{H1,ZEUS} have indeed
confirmed all the features of the predictions and have provided
unequivocal proof for the `resolved' processes.
\begin{figure}[hbt]
\vspace{7 cm}
\caption{Histogram of energy flow per event versus polar angle showing evidence
for spectator jet. Taken from first of the ref. \protect\cite{H1}}
\label{spectator}
\end{figure}
Fig. \ref{spectator} shows one of the experimental evidence.

Thus these observations have  provided a confirmation
(in addition to the DIS measurements) of the ideas about \f2gam\
and these experiments will continue to add to our knowledge of
the \glph , \vqisq .

\section{Beamstrahlung induced backgrounds at the next linear colldiers}
The above discussion explains in what sense one says that the photon has
hadronic structure. The discussion also shows that the  effects of
hadronic structure of the photon increase with increasing
photon energy. This  makes it clear that the
existence and study of the `resolved' processes at the current colliders is
necessary to understand the interactions of very high energy photons. One such
source of high energy photon is the phenomenon of `beamstrahlung' that will
exist at the next generation linear colliders.

\subsection{Beamstrahlung}
As explained in section 2, the next generation of $e^+e^-$ colliders will
operate at much higher luminosities  than those of the current ones. This is
partly necessiated by the reduction of the annihilation cross-section
with increasing energy. More importantly, due to the severe synchrotron
radiation losses at high energies, it is not possible to build a circular
\eplem\ collider beyond $\rts \geq 200 $ GeV. The higher energy colliders under
planning have to be therefore linear colliders, which operate in single
pass mode as opposed to the circular colliders where a bunch passes an
interaction point a number of times ({\it e.g.} at LEP this number is $\sim 10
^8$).  Hence to achieve the much higher luminosity that is needed the $e^+/e^-$
bunches will have to be extremely dense which in turn  causes the $e^-/e^+$
to see very
high electromagnetic fields due to the dense $e^+/e^-$ bunch. This causes
`bremsstrahlung' radiation . This radiation caused by the coherent
interactions of all $e^-/e^+$ with the $e^+/e^-$ bunch, is termed
`beamstrahlung' \cite{beam}. The  energy spectrum of the beamstrahlung photons
depends critically on the machine parameters and its calculation is an art in
itself. Fortunately approximate analytic expressions given by Chen \cite{chen}
are applicable for almost all the machine designs currently under
consideration \cite{fatdg}. The beamstrahlung  parameter $\Upsilon$ is
proportional to the effective magnetic field of the bunches and for Gaussian
beams the mean value of $\Upsilon$ is given by
\be
\Upsilon = {5 r_e^2 E N \over  6 \alpha_{em} \sigma_{z} m_e
                                (\sigma_x  + \sigma_y)} ,
\ee
where E is the beam energy, N is the number of electrons/positrons per bunch,
$\sigma_x$ and $\sigma_y$ are the transverse bunch dimensions, $r_e$ is the
classical electron radius. The expression shows that the beamstrahlung
parameter is larger for round bunches than for flat, ribbon--like  bunches.
For a given luminosity and bunch dimensions, the beamstrahlung can be reduced
by introducing more bunches. So beamstrahlung  can thus be controlled by
spatial/temporal shaping of the bunches.
There is also a suggestion \cite{telnov} to convert the \eplem\
linear colliders into \gamgam\ colliders by using the back-scattered lasers
from the $e^+/e^-$.  The photons in both these cases are `real' as opposed to
the `quasi-real' bremsstrahlung photons.
\begin{figure}[hbt]
\vspace{6 cm}
\caption{Photon spectra for $\protect\sqrt{s} = 500 $ GeV
for the different proposed
machine designs. WW is the Weizs\"acker Williams spectrum of the quasi--real
brmesstrahlung photons, and Laser shows the spectrum for the photons obtained
from a backscattered laser, taken from ref. \protect\cite{fatdg}}
\label{Beamstrah}
\end{figure}
Fig. \ref{Beamstrah} shows that the beamstrahlung spectra
are quite different for different machine designs all of which
correspond to roughly the same luminosity.

\subsection{Hadron production in \gamgam\ collisions and Beamstrahlung}
The net effect of beamstrahlung therefore is that associated with \eplem\
collision there is also a simultaneous \gamgam\ collision. The jet production
in
\gamgam\ collisions will have `direct' as well as the `1-res' and `2-res'
contributions as said before. The `resolved' contributions rise in importance
with increasing photon energies. This rise in $ \sigma(\gamgam \to  {\rm jets}
)$ has been experimentally confirmed in the laboratory as described in the
earlier sections. If one therefore now  extrapolates these calculations to
this situation, one finds a most unusual result.
A calculation \cite{fatdg} shows that the cross--sections for the
production of jets with very small $p_T$ in these \gamgam\ collisions  is very
large indeed.
{
\begin{table}[hbt]
\caption{Total `semi--hard' cross--section at the various
$\protect\sqrt{s} = 500$ GeV colliders for some machine designs and the total
number of events expected per effective bunch crossing for one parametrisation
of \f2gam\ , taken from \protect\cite{fatdg}.}
\begin{center}
\begin{tabular}{llll}
\hline
Collider & $\sigma^{\rm hard (DG)} (\mu b)$ & $ \sigma^{\rm soft} (\mu b) $ &
no. of events (DG)   \\
\hline
Tesla&0.016&0.041&0.004\\
D-D(wbb)&0.075&0.20&0.20\\
P-G & 0.48&0.51&24\\
Laser&1.9&0.25&0.49-95\\
\hline
\end{tabular}
\end{center}
\end{table}
}
Table 2 gives the integrated  semi--hard, inclusive cross--section
$\int_{\ptmin} {d \sigma (\gamgam\ \to {\rm jets)} \over dp_T}\;dp_T $ for
$\ptmin = 1.6$ GeV along with the `soft' cross--section that is expected on
the basis of the VMD picture mentioned earlier. The last column gives the
number of events containing small $p_T$ `jets' that will occur per `effective'
bunch crossings, i.e., bunch crossings which can not be distinguished from
each other. For the Laser machine the two numbers correspond to the TESLA and
Palmer-G design of the \eplem\ collider which is used to produce the \gamgam\
collider.  What this table tells us is that simultaneous to the  effective
\eplem\   event there  will be production of hadrons in the \gamgam\
collision and these hadrons will carry considerable energy  (e.g. for P-G
machine they will carry $ \simeq 24 *(1.6*2 + 2) \simeq 125 $ GeV ) which has
nothing to do with the \eplem\ collision and thus produce an underlying
event at an \eplem\ collider which is totally unheard of. Luckily, as the
table shows, the number of the underlying events depend very much on the
beamstrahlung and hence on the machine parameters. The machine designs can be
changed to reduce the beamstrahlung induced background. Eventhough the
inclusive cross-section that we have computed is not a measure of the total
\gamgam\ cross--section and also suffers from uncertainties
due to the poor knowledge of the gluon  content of the photon, it still gives a
measure of the `messiness' that would be caused by the underlying event at the
linacs. So this table underlies the need of studying  hadron production in
\gamgam\ collisions and controlling the beamstrahlung induced backgrounds at
the linacs \cite{chenpeskin}.

\subsection*{Acknowledgements}
I wish to thank the organisers of the symposium on 'Bose and 20 th
century Physics' for inviting  me to present this talk at the
symposium and give me an opportunity to pay my respects to this great
Indian Physicist. I also wish to thank M. Drees for an enjoyable  and
long standing  collaboration  during which some of the   results presented
in this talk were obtained. This work was in part supported by the
Council for Scientific and Industrial Research, Delhi under grant
no. 03(0745)/94/EMR-II.

\end{document}